\documentstyle[aps,prl,preprint,floats,epsfig]{revtex}

\begin{document}

\preprint{\tighten\vbox{\hbox{\hfil CLNS 97/1495}
                        \hbox{\hfil CLEO 97-15}
}}
             
\title{\bf Search for the Decay 
			$\tau^- \to 4\pi^- 3\pi^+ (\pi^0) \nu_\tau$}

\author{CLEO Collaboration}
\date{\today}

\maketitle
\tighten

\begin{abstract}
We have searched for the decay of the $\tau$ lepton into
seven charged particles and zero or one $\pi^0$.
The data used in the search were collected with the CLEO~II
detector at the Cornell Electron Storage Ring (CESR) and
correspond to an integrated luminosity of 4.61 $fb^{-1}$.
No evidence for a signal is found.
Assuming all the charged particles are pions, we set an upper
limit on the branching fraction, $B(\tau^- \to 4\pi^- 3\pi^+ (\pi^0) \nu_\tau)
< 2.4 \times 10^{-6}$ at the 90\% confidence level.
This limit represents a significant improvement over the previous limit.
\end{abstract}
\newpage

{
\renewcommand{\thefootnote}{\fnsymbol{footnote}}
\begin{center}
K.~W.~Edwards,$^{1}$
A.~Bellerive,$^{2}$ R.~Janicek,$^{2}$ D.~B.~MacFarlane,$^{2}$
P.~M.~Patel,$^{2}$
A.~J.~Sadoff,$^{3}$
R.~Ammar,$^{4}$ P.~Baringer,$^{4}$ A.~Bean,$^{4}$
D.~Besson,$^{4}$
% ...
\end{center}
                                                                             
\small
\begin{center}
$^{1}${Carleton University, Ottawa, Ontario, Canada K1S 5B6 \\
and the Institute of Particle Physics, Canada}\\
$^{2}${McGill University, Montr\'eal, Qu\'ebec, Canada H3A 2T8 \\
and the Institute of Particle Physics, Canada}\\
$^{3}${Ithaca College, Ithaca, New York 14850}\\
$^{4}${University of Kansas, Lawrence, Kansas 66045}\\
% ...
\end{center}
 
\setcounter{footnote}{0}
}
\newpage

The decay of the $\tau$ lepton into final states with seven or more pions
is of particular interest since it may provide a sensitive probe
of the $\nu_\tau$ mass due to the limited phase space.
There are several experimental upper limits
on the branching fractions (at the 90\% confidence level).
The HRS experiment~\cite{HRS} published an upper limit of
$B(\tau^- \to 4\pi^- 3\pi^+ \ge 0\ {\rm neutrals}\ \nu_\tau)
< 1.9 \times 10^{-4}$~\cite{conjugate}.
Recently, the OPAL experiment~\cite{OPAL} set an upper limit
of $B(\tau^- \to 4\pi^- 3\pi^+ (\pi^0) \nu_\tau)
< 1.4 \times 10^{-5}$.
For comparison, the upper limit on the branching fraction
for the decay $\tau^- \to 3\pi^- 2\pi^+ 2\pi^0 \nu_\tau$
as determined by the CLEO~II experiment~\cite{Gibaut}
is $1.1 \times 10^{-4}$.
In this paper, we present the result of a search for the
decay into seven charged particles and zero or one $\pi^0$
using the CLEO~II detector with the assumption that
all charged particles are pions.

The data used in this search were collected from $e^+e^-$
collisions at a center-of-mass energy ($\sqrt{s}$) of 10.6~GeV
with the CLEO II detector~\cite{Kubota} at the Cornell Electron
Storage Ring (CESR).
The total integrated luminosity of the sample is $4.61 fb^{-1}$,
corresponding to the production of $N_{\tau\tau}=4.21 \times 10^{6}$ 
$\tau$-pairs.
CLEO II is a general purpose spectrometer with excellent charged
particle and electromagnetic shower energy detection.
The momenta and specific ionization ($dE/dx$) of charged particles
are measured with three cylindrical drift chambers between 5 and
90 cm from the $e^+e^-$ interaction point that have a total of 67 layers.  
These are surrounded by a scintillation time-of-flight system
and a CsI(Tl) calorimeter with 7800 crystals.
These detector systems are installed inside a superconducting
solenoidal magnet (1.5 T), surrounded by an iron return yoke
instrumented with proportional tube chambers for muon identification.

The event selection criteria were designed to maintain a high
detection efficiency while suppressing the $\tau$ migration and hadronic
($e^+e^- \to q \bar q$) background.
The $\tau$ migration is primarily from the decays
$\tau^- \to 2\pi^- \pi^+2\pi^0\nu_\tau$ and
$\tau^- \to 3\pi^-2\pi^+ \pi^0\nu_\tau$, in which the $\pi^0$'s
decay via the Dalitz mechanism or via the $\gamma\gamma$ decay
channel with photon conversion at the beam pipe or drift
chamber walls.
Each $\tau^+\tau^-$ candidate event is required to contain eight
charged tracks with zero net charge.
The distance of closest approach of each track to the $e^+e^-$
interaction point must be less than 1~cm in the plane transverse to the beam 
axis and
10~cm along the beam axis; this requirement suppresses the
$\tau$ migration background from photon conversions.
Each track must have a momentum of at least
$0.02E_{beam}$ ($E_{beam} = \sqrt{s}/2$) and be in the
central region of the detector, ${|\cos \theta |<0.90}$, 
where $\theta$ is the polar angle with respect to the beam axis. 

The event is divided into two hemispheres using the plane
perpendicular to the thrust axis~\cite{thrust}, where the thrust axis
is calculated using both charged tracks and photons.
A photon candidate is defined as a calorimeter cluster with
a minimum energy of 60 MeV in the barrel region
($|\cos \theta |<0.80$) or 100 MeV in the
endcap region $(0.80 <|\cos \theta |<0.95)$.
The photon candidate must be isolated by at least 30~cm from
the projection of any charged track on the surface of the
calorimeter $\it and$ have either an
energy which is above 300~MeV or a lateral profile of energy
deposition consistent with that of a photon.
There must be one charged track in one hemisphere recoiling
against seven charged tracks in the other (1 vs.~7 topology)
with no more than two photons in the 1-prong hemisphere.
These requirements select the dominant one-charged-particle
decays of the $\tau$ lepton as tags, $\tau^- \to e^-\bar \nu_e \nu_\tau$,
$\mu^-\bar \nu_\mu \nu_\tau$, $\pi^- \nu_\tau$, $\rho^-\nu_\tau$, while
suppressing the hadronic background.
There is no photon multiplicity requirement in the 7-prong
hemisphere in order to minimize the dependence on the Monte
Carlo simulation (see below) of charged pions interacting
in the calorimeter that may mimic photon showers.
We also do not attempt to reconstruct the $\pi^0$ meson
in the decay $\tau^- \to 4\pi^- 3\pi^+ \pi^0 \nu_\tau$.
The migration background is further reduced by restricting
the number of electron candidates
in the 7-prong hemisphere to be no more than two~\cite{electron}.
An electron candidate is defined as a charged track with a
shower energy to momentum ratio in the range,
$0.85 < E/p < 1.1$, and, if available, a measured specific ionization
loss ($dE/dx$) consistent with that of an electron.

\begin{figure}[htbp]
\centerline{\epsfig{figure=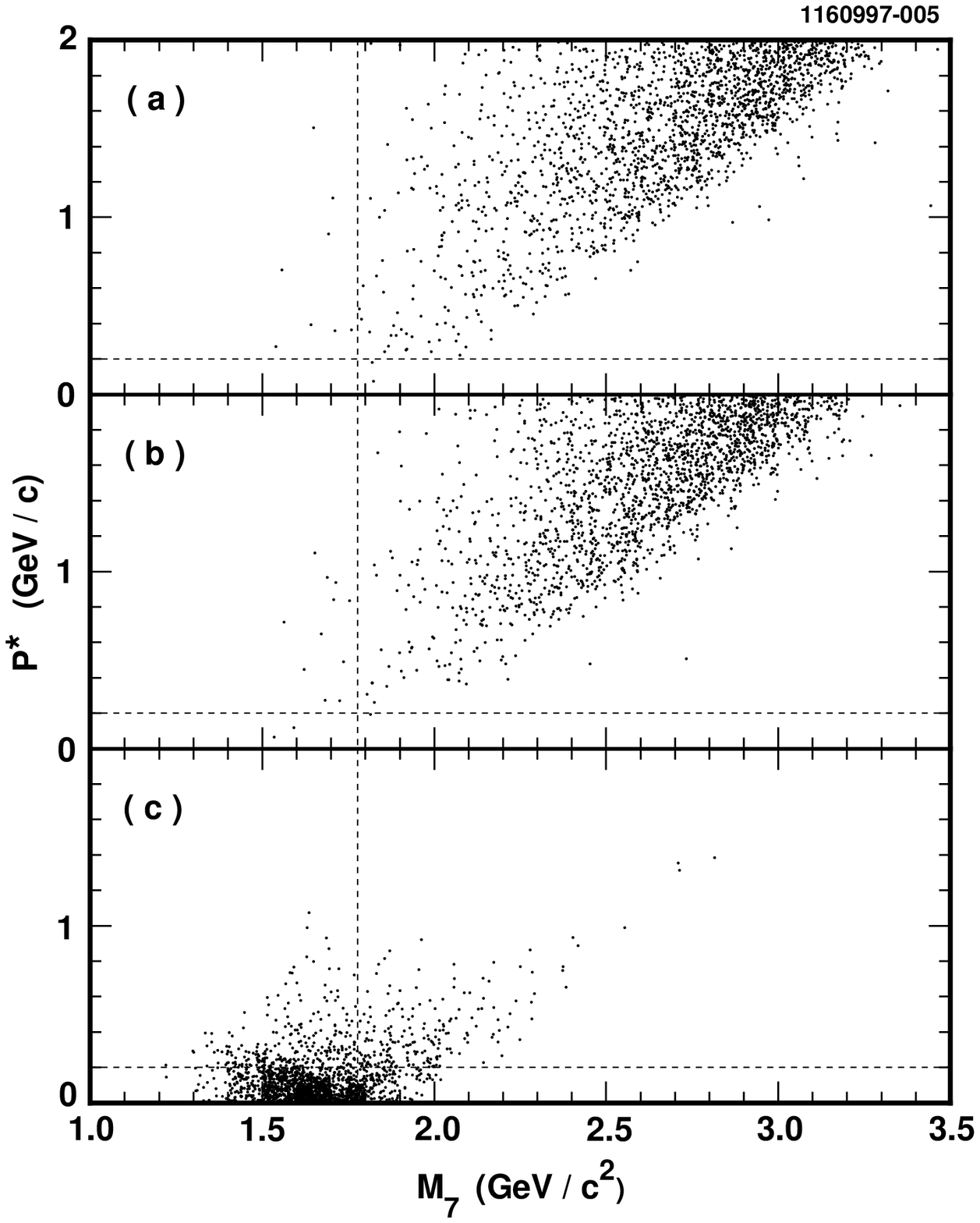,width=15cm}}
\
\caption{Center-of-mass momentum vs. invariant mass of the
7-prong hemisphere for the (a) data, (b) hadronic background,
and (c) signal Monte Carlo for $\tau^- \to 4\pi^- 3\pi^+ \nu_\tau$.
The hadronic background is obtained with a high mass tag,
$M_1 > 1.8 \rm\ GeV/c^2$.
The dashed lines indicate the values at which the 
respective cuts were imposed.}
\label{fig:doublehad}
\end{figure}

\begin{figure}[htbp]
\centerline{\epsfig{figure=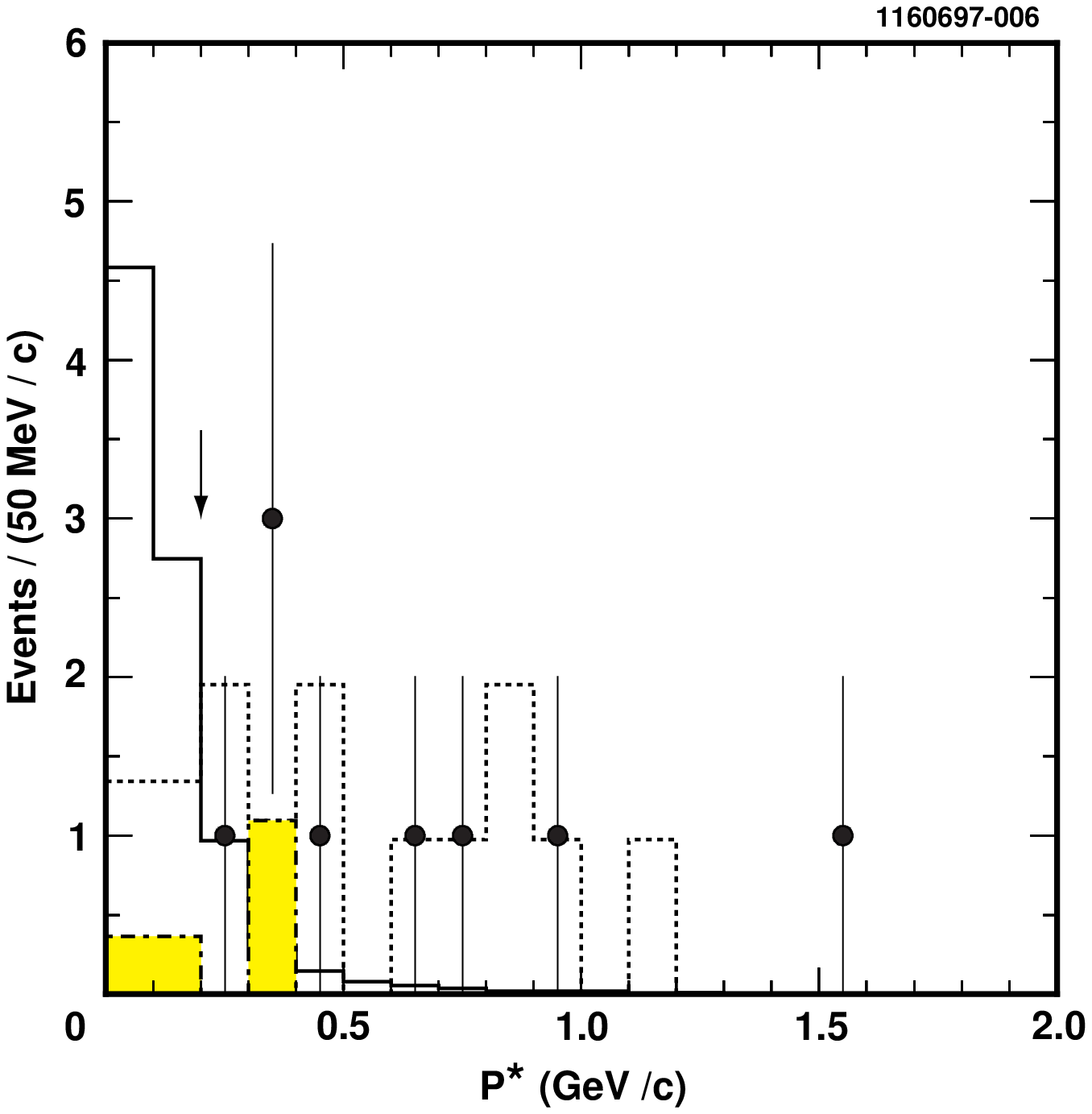,width=8cm}}
\caption{Center-of-mass momentum spectra of the 7-prong hemisphere
for the data, signal Monte Carlo for $\tau^- \to 4\pi^- 3\pi^+ \nu_\tau$
(solid), and background (dashed),
which includes the $\tau$ migration (shaded) and hadronic background.
The arrow indicates the value at which the cut was imposed.
The signal Monte Carlo is normalized to the number of events in the data.}
\label{fig:pstar}
\end{figure}

Two kinematic requirements are used to further reduce the hadronic 
background.
The total invariant mass of charged tracks and photons in each
hemisphere must be less than the $\tau$ mass
($M_1, M_7 < M_{\tau} = 1.777 \rm\ GeV/c^2$)~\cite{PDG}.  
The magnitude of the total momentum of the 7-prong hemisphere
in the $\tau$ rest frame, $P^*$, must be less than $0.2 \rm\ GeV/c$.
In calculating $P^*$, we assume the energy of the $\tau$ is the same
as that of the beam by ignoring initial state radiation and approximate
the $\tau$ direction by the direction of the 7-prong momentum vector. 
The $P^*$ requirement selects events with tau-like kinematics while
suppressing the hadronic background.
It also reduces the $\tau$ migration background from 
lower multiplicity decays in which the 7-prong jet momentum is not as good 
of an approximation of the $\tau$ direction. 
Figures~\ref{fig:doublehad}(a) and~\ref{fig:doublehad}(b) show
the $P^*$ vs.~$M_7$ distribution for the data and hadronic
background before the $P^*$ and $M_7$ requirements are imposed.
The hadronic sample is selected from the data using the criteria described
above, except that $M_1>1.8 \rm\ GeV/c^2$ and, to increase
statistics, there is no restriction on the photon multiplicity
in the 1-prong hemisphere.
The hadronic background shows a cluster of events in the
region of large $P^*$ vs.~$M_7$.
However, the Monte Carlo (see below) predicts an enhancement of
events with low $P^*$ and $M_7$ for the signal decay
$\tau^- \to 4\pi^- 3\pi^+ \nu_\tau$ as
shown in Fig.~\ref{fig:doublehad}(c).
The $P^*$ distribution for the data events with $M_7 < M_\tau$
is shown in Fig.~\ref{fig:pstar}.
It is evident from both Fig.~\ref{fig:doublehad}(a) and Fig.~\ref{fig:pstar}
that no events satisfy the selection criteria described above.

The detection efficiencies ($\epsilon$) for the signal decays
are estimated by Monte Carlo simulation.
The KORALB/TAUOLA generator is used to create $\tau$ pairs according to
the standard electroweak theory, including $\alpha^3$ radiative
corrections~\cite{Jadach}.
The decays $\tau^- \to 4\pi^- 3\pi^+ \nu_\tau$ and
$4\pi^- 3\pi^+ \pi^0 \nu_\tau$ are
modeled using phase space with a V-A interaction.
The GEANT program~\cite{Brun} is used to simulate the detector response.
The estimated detection efficiencies are given in Table~\ref{table:results}.
The two efficiencies are comparable, as expected, since the same selection
criteria were imposed on these two kinematically similar decays.
As a test of the validity of the analysis, we compare the zero
events observed with the expected number of events from the
$\tau$ migration and hadronic background.
The migration background is determined using the Monte Carlo technique.
The hadronic background is empirically estimated from
Fig.~\ref{fig:doublehad}(b) with the assumption that
$M_1$ and $M_7$ are not strongly correlated;
this is evident from the similarity of Figs.~\ref{fig:doublehad}(a)
and \ref{fig:doublehad}(b) in the region of large $P^*$ vs.~$M_7$.
The number of hadronic events with $M_7 < M_\tau$ is determined by
normalizing the number of events in Fig.~\ref{fig:doublehad}(b)
with $M_7 > 1.8 \rm\ GeV/c^2$ to that in Fig.~\ref{fig:doublehad}(a).
The estimated background is summarized in Table~\ref{table:results}.
The observation of zero events is consistent with the background expectation.

\begin{table}[b]
\begin{center}
\caption[]{Summary of signal, background, efficiency, and branching fraction
(at the $90\%$ confidence level).  All errors are statistical.}
\vspace{0.1in}
\label{table:results}
\begin{tabular}{lc}
Data		                                	&	0                      \\ \hline
$\tau$ migration 		                    &	$0.88 \pm\ 0.23$       \\ \hline
${q\bar q}$ background	                &	$1.95 \pm\ 1.40$       \\ \hline
$4\pi^-3\pi^+     $ efficiency (\%)	   &	$15.7 \pm\ 0.2$        \\ \hline
$4\pi^-3\pi^+\pi^0$ efficiency (\%)  	 &	$15.9 \pm\ 0.3$        \\ \hline\hline
$B(\tau^- \to 4\pi^-3\pi^+(\pi^0)\nu_\tau)$	&	$< 2.38 \times 10^{-6}$ \\ 
\end{tabular}
\end{center}
\end{table}

The upper limit of the branching fraction is determined from the upper
limit on the number of candidate events $\lambda$ by
$$B(\tau^- \to 4\pi^- 3\pi^+ (\pi^0) \nu_\tau)
= \frac {\lambda}{2\epsilon B_{tag}N_{\tau\tau}}\,$$
\noindent where $B_{tag} = (73.0 \pm 0.3)\%$ is the sum of branching
fractions of the tags~\cite{PDG}.
Since the efficiencies for $4\pi^- 3\pi^+$ and $4\pi^- 3\pi^+\pi^0$
are the same within 1\%, we choose the lower efficiency for
the former decay to derive a more conservative upper limit.
Using $\lambda = 2.30$ events for the zero candidate events
observed, the result is
$ B(\tau^- \to 4\pi^- 3\pi^+ (\pi^0) \nu_\tau) < 2.38 \times 10^{-6}$
at the 90\% confidence level.

The systematic error contains contributions from several sources.
These include the 1\% uncertainty in the luminosity, 1\% uncertainty in
the $\tau^+\tau^-$ production cross-section, 12\% uncertainty in the tracking
efficiency, 0.3\% uncertainty in the branching fraction of the tag
as well as the 1\% statistical error in the detection efficiency
due to limited Monte Carlo statistics.
The systematic error in the tracking efficiency is estimated from
a study of the charged particle multiplicity distribution of
hadronic events.
Since the tracks from 7-prong $\tau$ decays are more
collimated, we also study the ability of the Monte Carlo to
simulate reconstruction of collimated tracks by comparing the
minimum opening angle distribution of like-sign tracks in
5-prong $\tau$ decays from the data with that of the Monte Carlo.
The reproduction by the Monte Carlo is quite satisfactory.
The total systematic error is calculated by adding all errors
in quadrature.
The final result at the 90\% confidence level is
$$ B(\tau^- \to 4\pi^- 3\pi^+ (\pi^0) \nu_\tau)  < 2.4 \times 10^{-6},$$
where Gaussian statistics were used to include the systematic 
error~\cite{Cousins}.

In conclusion, we find no evidence for the decay
$\tau^- \to 4\pi^- 3\pi^+ (\pi^0) \nu_\tau$ and set
an upper limit on the decay branching fraction.
The upper limit is significantly more stringent than those
of the previous experiments~\cite{HRS,OPAL}.

We gratefully acknowledge the effort of the CESR staff in providing us with
excellent luminosity and running conditions.
This work was supported by 
the National Science Foundation,
the U.S. Department of Energy,
the Heisenberg Foundation,  
the Alexander von Humboldt Stiftung,
Research Corporation,
the Natural Sciences and Engineering Research Council of Canada,
and the A.P. Sloan Foundation.



\newpage

\begin{thebibliography}{99}

\bibitem{HRS} B.G. Bylsma {\it et al.}, Phys. Rev. {\bf D35}, 2269 (1987).

\bibitem{conjugate} In this paper charge conjugate states are implied.

\bibitem{OPAL} K.~Ackerstaff {\it et al.}, CERN Report No. CERN-PPE/97-049,
        1997 (submitted to Phys. Lett. B).

\bibitem{Gibaut} D.~Gibaut {\it et al.},
        Phys.~Rev. Lett.~{\bf 73}, 934 (1994).

\bibitem{Kubota} Y. Kubota {\it et al.}, Nucl. Instrum. Methods A{\bf 320}, 66 (1992).

\bibitem{thrust} E.~Farhi, Phys. Rev. Lett.~{\bf 39}, 1587 (1977).

\bibitem{electron} We allow up to two electron candidates in the 7-prong hemisphere
                   in order to minimize the dependence on the Monte
                   Carlo simulation of charged pions interacting
                   in the calorimeter that may mimic electrons.

\bibitem{PDG} R. Barnett {\it et al.}, Review of Particle Properties,
        Phys. Rev. {\bf D54}, 1 (1996).

\bibitem{Jadach} S. Jadach and Z. Was, Comput. Phys. Commun. {\bf 36}, 191 (1985);
        {\bf 64}, 267 (1991); S.~Jadach, J.H. Kuhn, and Z. Was, {\it ibid}. 
        {\bf 64}, 275 (1991).

\bibitem{Brun} R. Brun {\it et al.}, CERN Report No. CERN-DD/EE/84-1, 1987 (unpublished).

\bibitem{Cousins} R.D. Cousins and V.L. Highland, Nucl. Inst. and Meth. A{\bf 320}, 331 (1992).

\end{thebibliography}
\end{document}